\documentstyle[12pt]{article}

\def\Journal#1#2#3#4{{#1} {\bf #2}, #3 (#4)}

\def\AP{\em Ann. Phys.}

\def\NPB{{\em Nucl. Phys.} B}
\def\PLB{{\em Phys. Lett.}  B}
\def\PRL{\em Phys. Rev. Lett.}
\def\PRD{{\em Phys. Rev.} D}

\def\PTPS{{\em Prog. Theor. Phys. Suppl.}}

\font\tenrm=cmr10
\font\sevenrm=cmr7

\newfam\BMath
\font\BMathL=cmmib10
\font\BMathl=cmmib7
\font\BMathm=cmmib5
\textfont\BMath=\BMathL \scriptfont\BMath=\BMathl
\scriptscriptfont\BMath=\BMathm
\def\A{{\fam\BMath A}}

\def\K{{\fam\BMath k}}
\def\X{{\fam\BMath x}}
\def\Y{{\fam\BMath y}}

\def\a{\alpha}
\def\b{\beta}
\def\d{\delta}
\def\f{\phi}
\def\g{\gamma}
\def\j{\psi}
\def\q{\theta}
\def\o{\omega}
\def\ca{{\cal A}}
\def\ce{{\cal E}}
\def\cl{{\cal L}}
\def\ch{{\cal H}}

\def\intks{\int \frac{d^3 \K}{(2\pi)^3}}
\def\intxs{\int d^3 \X}
\def\exp{\mbox{\rm exp}}

\def\ra{\rightarrow}
\def\Del{\mbox{\boldmath $\partial$}}

\def\srm#1{\mbox{\sevenrm #1}}

\def\eff{\srm {eff}}

\def\be{\begin{equation}}
\def\ee{\end{equation}}
\def\bea{\begin{eqnarray}}
\def\eea{\end{eqnarray}}
\def\eref#1{Eq.~(\ref{#1})}

\newcommand{\ncom}{\newcommand}
\ncom{\vo}[1]{{\fam\BMath #1}}
\ncom{\vt}[2]{({\fam\BMath #1}-{\fam\BMath #2})}
\ncom{\lan}{\langle}
\ncom{\ran}{\rangle}
\ncom\nonum{\nonumber \\}
\ncom\fx{\!\!\!\!}

\ncom{\half}{{1\over 2}}
\ncom{\third}{{1\over 3}}
\ncom{\fourth}{{1\over 4}}
\ncom{\fifth}{{1\over 5}}
\ncom{\sixth}{{1\over 6}}

\begin{document}

\rightline{\tenrm LPTHE-Orsay 95/66}

\vskip 1cm
\centerline{\bf DEBYE SCREENING, KMS AND IMAGINARY-TIME}
\centerline{\bf FORMALISM OF TEMPORAL AXIAL GAUGE
\footnote{Talk presented at Thermo95, Dalian, China}}

\vskip 0.5cm
\centerline{S.M.H. WONG }

\vskip 0.5cm
\centerline{\footnote{Laboratoire associ\'e au Centre National
de la Recherche Scientifique}\it LPTHE, Universit\'e de Paris XI,
B\^atiment 211,\\ F-91405 Orsay, France}

\vskip 1.0cm
{\tenrm We argue that in QCD, the Debye mass requires not only a
mathematical definition but also a physical one and temporal axial
gauge could provide for a physical screening potential for this
purpose. Unfortunately, this gauge is spoiled by problem of energy
conservation rather than the well known divergence due to the double
pole in the longitudinal propagator. We also show that KMS
condition is violated in this gauge and is therefore not universally
true.}

\section{Why Imaginary-Time and Why Temporal Axial Gauge?}

Since the beginning of the nineties, it is known that at finite
temperature, in order to do true perturbative calculations
order by order in the coupling constant, it is necessary to perform
Braaten-Pisarski resummation\cite{bra&pis} and use effective
propagators and vertices in place of the bare quantities. This
resummation requires the hard thermal loops\cite{fre&tay} of
the N-point functions. These are essentially the leading terms of
the one-loop N-point functions. In real-time, there is the
doubling of the degrees of freedom, so in principle, one
will have to work out all the components of each N-point hard
thermal loops before one can do resummation. This is, however, not
necessary in imaginary-time. There is only one hard thermal loop for
each N-point function so imaginary-time is comparatively simple to
work in. Furthermore, when considering static problem like calculating
the Debye mass at the next to leading order, there is the
simplification of it is only necessary to keep the zero mode
\cite{reb1,pei&won} in the Matsubara frequency sum.
Other modes are irrelevant at this order.

Temporal axial gauge (TAG) offers one the chance to work in an
``abelianized'' gauge field theory as long as one restricts oneself to
the static chromoelectric sector. Therefore in this gauge, one can
say with certainty, that the gauge invariant static quark-antiquark
potential in a hot medium is directly related to the longitudinal
gluon propagator alone. In other gauges, 3-point and/or
4-point function cannot be easily excluded. This will be shown more
explicitly in Sec. \ref{sec:debye}. TAG has another feature which is
the absence of ghost, that means one cannot include the whole Hilbert
space in the partition function which leads to the violation of KMS
boundary condition. This will be shown in Sec. \ref{sec:kms}.

Having mentioned the physical reasons and advantages of working in TAG,
one must not forget the disadvantage. It is well known that the
longitudinal propagator has a troublesome double pole $1/k_0^2$ at T=0.
To handle this pole, some prescription is required to displace it away
from the real energy axis. At finite temperature, if one simply turns
the T=0 propagator into the finite T propagator by giving it discret
imaginary energy, one will be facing immediately a divergence at zero
energy. This problem has traditionally been dealt with by simply
dropping the infinity\cite{hei&kaj&toi}. It is found to be correct at
leading order but at higher order it is almost certainly not correct.
It is simple to understand why dropping the divergence will not affect
the leading order result. Since the leading terms are essentially the
hard thermal loops which do not get any contribution from the soft
zero mode and the divergence is precisely coming from this mode.
In the following, it will be shown that in fact this problem of the
double pole does not exist at finite T in the imaginary-time formalism,
however, the hope of using a trouble free imaginary-time formalism of
TAG to study physical problems can still not be fulfilled.

\section{Debye Screening}
\label{sec:debye}

\subsection{Debye Screening in TAG}

The potential of a charge $Q_1$ at ${\vo x}_1$ in the
presence of another charge $Q_2$ at ${\vo x}_2$ is given by
\cite{kaj&kap}
\be V(\vo r) = {1\over 2}\sum_a \intxs \, \left (
                \ce_1^{a \;\eff} (\vo x)
                \cdot \ce_2^a (\vo x) +
                \ce_2^{a \;\eff} (\vo x)
                \cdot \ce_1^a (\vo x) \right )
\label{eq:potenteqEE}
\ee
where the effective field $\ce_1^{a \;\eff}$ created by
the non-abelian charge $Q_1^a$ is the sum of the applied field
$\ce_1^a$ and the induced field
$\delta \lan\vo E_1^a \ran$:
\be \ce_1^{a \;\eff}(\vo x) =
    \ce_1^a(\vo x) + \delta \lan \vo E_1^a \ran
\ee
$\ce_1^a$ is a solution of the Gauss' law:
\be \nabla \cdot \ce^a -g f^{abc}
     \ce^b \cdot \ca^c = Q_1^a
     \delta^3(\vo x -\vo x_1)  \; ,
\label{eq:gauss}
\ee
where $\ca$ is the vector potential associated with $\ce$.
The potential \eref{eq:potenteqEE} is manifestly gauge invariant
and is therefore physical.

In TAG, the electric field is linear in the vector potential
\be E_i^a(\vo x, t) =-\partial_0 A^a_i(\vo x, t) \; ,
\label{eq:EinTAG}
\ee
and in a static situation, $A$'s depend linearly in time so
Gauss' law becomes abelian. It is now simple to solve
and the solution is
\be \ce_i^a(\vo x) = -i Q^a_1 \intks
     e^{i\K \cdot (\X-\X_1)} {k_i \over {\K}^2} \; .
\label{eq:gausssolut}
\ee
We stress that in gauges such as covariant or Coulomb gauge, Gauss'
law is not abelian and is therefore not trivial to solve\cite{pei&won}.
The form \eref{eq:gausssolut} plays an important role in determining
to which N-point function the physical screening potential
\eref{eq:potenteqEE} is related.

The coupling Hamiltonian which couples the external applied field
$\ce^a_1$ to the field in the medium $\vo E^a_1$ is
\be  H^{\srm{ext}} (t)= \intxs \,\vo E^a_1(\vo x, t)
                      \cdot \ce_1^a(\vo x) \; .
\ee
So from linear response theory, the induced field is
\bea \lan E^a_i(\vo x, t) \ran &=& i\int^t_{-\infty}
     dt'\lan [H^{\srm{ext}} (t'), E^a_i(\vo x, t)] \ran  \nonumber \\
     &=& i\int^t_{-\infty} dt' \intxs' \ce_j^b(\vo x')
     \lan [E^b_j(\vo x', t'), E^a_i(\vo x, t)] \ran \; .
\label{inducedfield}
\eea
The correlator $\lan [E,E]\ran$ can be written in terms of the
retarded gluon propagator because of \eref{eq:EinTAG},
so the effective field in the plasma is
\be \ce^{a \;\eff}_i (\X) = \intks e^{i\K \cdot \X} \ce^b_j (\K)
    \lim_{\o\ra 0} [\o^2 D^{ab\; R}_{ij} (\o, \K)] \; .
\ee
After putting everything back into \eref{eq:potenteqEE}, we see that
the retarded propagator is now contracted between two $\ce^a$'s which
project out the longitudinal component of the propagator. So in TAG,
the screening potential is directly related only to the longitudinal
gluon propagator. If \eref{eq:gausssolut} is not of such a form,
then this would not be true which is the case in other gauges.

\subsection{Definition of the Debye Mass}

In this section, we would like to argue that although the new
definition of the Debye Mass proposed by Rebhan defined at the pole
of the longitudinal propagator
\be
    m^2 = \lim_{k^2\ra -m^2} \Pi_{00}(0,k)
\label{eq:massdef}
\ee
is significantly improved over the old definition
\be
    m^2 = \lim_{k^2\ra 0} \Pi_{00}(0,k)
\ee
in the sense that it is self-consistent and also gauge\cite{kob&kun&reb}
and renormalization group invariant\cite{reb2}. It still remains only a
good mathematical definition. It is necessary but not sufficient.
Because to be certain that this mass is indeed the Debye mass,
one will have to first find a physical static quark-antiquark
potential which has an exponential form at large spatial separation
and then the inverse screening length will have to be exactly given by
\eref{eq:massdef}.

As explained in the previous section, it is not clear that one can
look for screening behaviour using only the 2-point function
in gauges other than TAG. For example, in covariant and Coulomb
gauge, the screening function\cite{reb1} based on the longitudinal
gluon propagator is not gauge invariant and is therefore not physical.
So even if it behaves exponentially at large distance, one still cannot
say that the screening length of this function is the inverse of the
Debye mass. This is further complicated by the need to introduce by
hand the magnetic mass both to remove infrared divergence and to ensure
gauge invariance\cite{kob&kun&reb}. For a discussion on the consistency
of this, we refer to the paper of Blaizot and Iancu\cite{bla&ian}.

\section{KMS is not Universal}
\label{sec:kms}

 From quantum mechanics, the probability amplitude for evolving from a
state $q$ at time $t$ to a state $q'$ at time $t'$ can be written as
\be \lan q'\; t' | q\; t \ran = \int [dq] \; \exp \{ -i \int^{t'}_t dt L \} \;
{}.
\ee
If one applies this to the partition function for, say scalar field theory,
\bea Z_\f &=& \sum_\f \; \lan \f | \; \exp (-\b H_\f) \; | \f \ran  \nonum
           &=& \int_{\srm {periodic}} [d\f] \;
           \exp \{ -i \int^{-i\b}_0 dt L_\f \}    \; .
\eea
The interpretation is that one starts from a state $\f$ at $t=0$
and evolves back through a time $-i\b$ to $\f$, so $\f$ has to
be periodic in $-i\b$. In the case of free gauge fields in TAG,
since only transverse (physical) states are included in the thermal
average, the partition function is
\bea Z_{A} &=& \sum_T \; \lan T| \; \exp (-\b H_A) |T \ran  \nonum
            &=& \int_{\srm {periodic}} [dA_T] \;
            \exp \{ -i \int^{-i\b}_0 dt L_{A_T} \}   \; .
\eea
The longitudinal field part of the Hamiltonian has only the vacuum
to act on so there is no path integral for the longitudinal field.
We see that the transverse field must be periodic in time but
the longitudinal field is not required to be so.

Periodic field implies KMS boundary condition therefore the transverse
propagator is periodic. Whereas the longitudinal field is not periodic,
moreover, there is no trace identity tr($AB$)=tr($BA$) due to the
unphysical part of the Hilbert space is excluded so the longitudinal
propagator does not satisfy KMS. This feature is also true in real-time.
Therefore KMS does not hold universally as is widely assumed.

\section{The Double Pole $1/k_0^2$ Problem does not exist at Finite T}

We start by setting $A_0=0$ which we can do if there is no divergence
due to the $1/k_0^2$. We will assume this and check that this is the
case below. Continuing to work in the free field case, the lagrangian is
\be \cl = \half (\dot \A^2 + \A_T \Del^2 \A_T )
\ee
so the equation of motion of the longitudinal field is $\ddot \A_L=0$
and it is not a wave equation.

In order to quantize $\A_L$, we write down a general form for it
\be \A_L(\X) = \intks e^{i\K \cdot \X} [\a(\K) +\b(\K)\, t]\, \hat \K / k
    \; .
\ee
$\A_L$ is Hermitian so we can rewrite this as
\be \A_L(\X) =\intks \Big \{e^{i\K\cdot\X} [\a(\K) +\b(\K)\, t]
              +e^{-i\K\cdot\X} [\a^\dagger(\K) +\b^\dagger(\K)\, t]
              \Big \} \q(k_3) \hat \K / k \; .
\label{eq:a_l}
\ee
$k_3$ in the theta function is of course arbitrary. One can equally choose
$k_1$ or $k_2$. The commutation relations for the operators, $\a$, $\b$,
$\a^\dagger$ and $\b^\dagger$ are to be fixed by the canonical
commutation relations.

Because of the presence of the theta function, canonical commutation
relations such as $[\A_L(\X), \A_L(\Y)]=0$ and
$[\dot \A_L(\X), \dot \A_L(\Y)]=0$ cannot be satisfied trivially
with the usual type of commutation relations like
$[\a(\K),\a^\dagger(\K')]= 2 k \; (2\pi)^3 \d(\K-\K')$ and a similar
one for $\b$ and $\b^\dagger$. Instead, one is obliged to choose
\bea [\a(\K), \a^\dagger(\K')] \fx &=& \fx 0 \; , \mbox{\hskip 1.5cm}
     [\b (\K), \b^\dagger(\K')]=0   \; , \nonum
     {[}\a(\K), \b^\dagger(\K')] \fx &=& \fx i\, (2\pi)^3 \K^2 \d(\K-\K')
\label{eq:com_rel}
\eea
and the Hermitian conjugate of the last relation above.

With \eref{eq:a_l} and \eref{eq:com_rel}, we can now work out the
longitudinal gluon propagator in configuration space. The momentum space
representation can be obtained by Fourier transform and the full time
range, i.e. from $-i\b$ to $i\b$, must be used in order to go to
energy space due to the lack of periodicity. Since the physical
states are the $|T\ran$'s so the longitudinal gluon propagator is
a T=0 propagator. It is not heated at the lowest order and
it has the form\cite{jam&lan,wong}
\be D^L_{ij}(t,\K)= {{k_i k_j} \over \K^2} T
    \sum_{k_0 \; \mbox{\scriptsize odd, even} }
    D^L_{k_0}(\K) e^{ik_0 t} \; .
\ee
The momentum space form of the zero mode component is
$D^L_{k_0=0}(\K)=1/{4T^2}$. So we see that the double pole $1/k_0^2$, in
fact, does not exist in agreement with the assumption we made at the
beginning of this section.

\section{Energy Conservation}
\label{sec:ene_con}

In the last section, we see that $D^L$ has both even and odd modes
which are commonly named erroneously bosonic and fermionic modes
respectively. The presence of both types of mode in a propagator
is actually problematic, if one recalls how energy conservation is
ensured in the usual case. In any interaction, it is the time
integration at each vertex which gives the important energy conserving
delta function. Consider any 3-point interaction in imaginary-time,
the time integration is of the form
\be \int^{-i\b}_0 dt \; e^{i(k_{01}+k_{02}+k_{03}) t} = -i\b \;
    \d_{k_{01}+k_{02}+k_{03},0}
\ee
provided the sum of the energies $k_{0i}, i=1,2,3$ of the three incoming
particles is an even multiple of $2\pi i T$. This is always the case in
non-gauge theories. In a gauge theory, eg. QED in TAG, this is not
always the case because the sum of the energies of the interaction
$e \bar \j \g^i \j A_{Li}$ can now be both even and odd. So the energy
conservation mechanism that one usually has is broken in this
case\cite{wong}.

If one looks at this from another angle, one can write down a thermal
N-point function in terms of thermal average over physical states of
N Heisenberg fields. Re-expressing everything in the interaction
picture and using the properties of the interaction picture fields
to introduce a time shift, say $\d$, to every one of the N fields.
The resulting expression\cite{wong} differs from the initial
expression in the interaction picture by only the time shift $\d$ in
the N fields plus $\exp(\pm i\b H)$ on either side of the kernel
of the matrix elements, acting on the enclosing $\lan phys|$ and
$|phys \ran$. Since physical states can be rewritten as energy
eigenstates of $H$, so the exponential operators become c-numbers
and cancel each other. Therefore any N-point function satisfies
\be \Gamma^N (t_1,t_2,\cdots,t_N)=\Gamma^N (t_1+\d,t_2+\d,\cdots,t_N+\d)
\ee
and since time-translation invariance implies energy conservation so
the latter still somehow seems to hold, despite the fact that the
simplest energy conservation mechanism is broken.

\section{Indefinite Metric Field Theory}

In choosing TAG and setting $A_0=0$, $\A_L$ depends linearly on time
and on four operators which satisfy \eref{eq:com_rel}.
Using these, one can construct states of the form
\be |\j\ran = \prod_\K (\a^\dagger(\K))^{m_\K} (\b^\dagger(\K))^{n_\K}
    |0\ran \; ,
\label{eq:gensta}
\ee
where $m_\K$ and $n_\K$ are both integers.
The simplest states are
\bea |\a\ran \fx &=& \fx \a^\dagger(\K)|0\ran \; ,  \\
     |\b\ran \fx &=& \fx \b^\dagger(\K)|0\ran \; ,  \\
     |\a,\b\ran \fx &=& \fx \a^\dagger(\K) \b^\dagger(\K)|0\ran \; ,
\eea
of which, the first two have zero norms. While the second is an eigenstate
of the free longitudinal Hamiltonian $H^L_0$ of zero eigenvalue, the other
two are not. Furthermore, they cannot be made into eigenstates of $H^L_0$
by superposition. The Hilbert space of $H^L_0$ is in fact spanned
by an infinite number of null and non-null states given by
\eref{eq:gensta}, not all of which are eigenstates of $H^L_0$. In the
language of indefinite metric field theory\cite{nak}, this Hilbert space
is spanned instead by the generalized eigenstates of $H^L_0$.

The generalized eigenstates $|\,\o\ran$ of a Hamiltonian $\ch$ and their
corresponding generalized eigenvalues $\o$ are defined by
\bea  (\ch - \o)^p |\,\o\ran &=& 0  \mbox{\hskip 1cm for $p\geq n$,}
      \nonum
                          &\neq& 0 \mbox{\hskip 1cm otherwise,}
\eea
for some integer n. In our present case, the generalized eigenvalues
are all zeros and the $n$ for the first and third states above is 2.

Generalized eigenstates are of course not the same as eigenstates
if $n\neq 1$, so the argument at the end of Sec. \ref{sec:ene_con}
unfortunately does not work in the final step. So although the problem
of the double pole no longer exists, we are facing a new obstacle
of energy conservation forced upon by the Hamiltonian formulation.

\section*{Acknowledgments}

The author thanks the organizers and the people of Dalian
University of Technology for all their efforts and also H. Matsumoto
for helpful discussions. The author acknowledges financial
support from the Leverhulme Trust.

\end{document}